\documentclass[prl,twocolumn,aps,longbibliography,superscriptaddress,notoc]{revtex4-1}
\usepackage{bm}
\usepackage{graphicx}
\usepackage{amssymb}
\usepackage{amsmath}
\usepackage{eufrak}
\usepackage{color}
\usepackage[utf8]{inputenc}
\usepackage{ulem}
\usepackage[unicode=true,colorlinks=true,citecolor=blue,urlcolor=blue]{hyperref}

\let\ifr\i
\renewcommand{\i}{{\rm i}}
\renewcommand{\d}{\mathrm d}
\renewcommand{\emph}{\textit}
\newcommand{\braket}[1]{\left\langle #1 \right\rangle}

\newcommand{\enquote}{}

\newcommand{\nix}[1]{}
\let\oldsec\section
\renewcommand{\section}[1]{\textit{#1}---}

\begin{document}

\title{Dynamic polarization of electron spins\\
interacting with nuclei
in semiconductor nanostructures}

\author{D.~S.~Smirnov}
\email[Electronic address: ]{smirnov@mail.ioffe.ru}
\affiliation{Ioffe Institute, Russian Academy of Sciences, 194021 St. Petersburg, Russia}
\author{T.~S.~Shamirzaev}
\email[Electronic address: ]{sha\_tim@mail.ru}
\affiliation{Rzhanov Institute of Semiconductor Physics, Siberian Branch of the Russian Academy of Sciences, 630090 Novosibirsk, Russia}
\affiliation{Ural Federal University, 620002 Yekaterinburg, Russia}
\author{D.~R.~Yakovlev}
\affiliation{Experimentelle Physik 2, Technische Universit\"at Dortmund, 44221 Dortmund, Germany}
\affiliation{Ioffe Institute, Russian Academy of Sciences, 194021 St. Petersburg, Russia}
\author{M.~Bayer}
\affiliation{Experimentelle Physik 2, Technische Universit\"at Dortmund, 44221 Dortmund, Germany}
\affiliation{Ioffe Institute, Russian Academy of Sciences, 194021 St. Petersburg, Russia}

\begin{abstract}
We suggest a new spin orientation mechanism for localized electrons:  \textit{dynamic electron spin polarization provided by nuclear spin fluctuations}. The angular momentum for the electrons is gained from the nuclear spin system via the hyperfine interaction in a weak magnetic field. For this the sample is illuminated by an unpolarized light, which directly polarizes neither the electrons nor the nuclei. We predict, that for the electrons bound in localized excitons $100$\% spin polarization can be reached in longitudinal magnetic fields of a few millitesla. The proof of principle experiment is performed on momentum-indirect excitons in (In,Al)As/AlAs quantum dots, where in a magnetic field of $17$~mT  the electron spin polarization of $30$\% is measured.
\end{abstract}

\maketitle{}

As the quantum computation era is believed to approach~\cite{arute2019quantum}, the investigations  of the underlying physics drastically intensify. Particular attention is focused on the spin dynamics of  localized electrons in semiconductor nanostructures~\cite{book_Glazov}, which is at the heart of various quantum computation and quantum cryptography schemes~\cite{PhysRevLett.89.147902,delteil2016generation}. Ultrafast optical orientation~\cite{PhysRevLett.94.047402,PhysRevLett.99.097401,gerardot08}, manipulation~\cite{greilich06,press08,Greilich2009} and readout~\cite{berezovsky2006nondestructive,atature07,Arnold2015} were already demonstrated for the single electrons confined in quantum dots (QDs).

There are two main approaches to spin orientation in nanostructures: optical spin orientation~\cite{OptOr} and thermal spin polarization in magnetic field. The first one does not require the external magnetic field as it is based on the transfer of the angular momentum from circularly polarized photons to electrons through the spin-orbit interaction. The second approach does not require optical excitation. It needs lowering of the lattice temperature, so that the thermal energy becomes smaller than the electron Zeeman splitting.

In this Letter, we suggest another approach, which we call the dynamic electron spin orientation. Generally, this concept applies to localized electrons exposed to unpolarized optical excitation. It requires: (i) a fine structure splitting of a photogenerated electron-hole pair with different lifetimes of the individual levels and (ii) that these levels are mixed by the random Overhauser field, i.e. the local magnetic field caused by the fluctuations of the host lattice nuclear spins. In III-V and II-VI semiconductors, the hyperfine interaction with nuclei is most pronounced for electrons, while for holes it is an order of magnitude weaker~\cite{book_Glazov}. The typical value of the random Overhauser field scales with the localization volume $V$ as $1/\sqrt{V}$. Therefore, for delocalized or weakly bound states the hyperfine interaction is negligible, while for electrons in QDs it plays the main role in the spin dynamics. We predict, that application of a weak external magnetic field of the order of the random Overhauser field (a few mT) induces an electron spin polarization, that can reach 100\%.

In contrast to the optical spin orientation, the proposed mechanism does not require circular polarization of the optical excitation. In contrast to the thermal spin polarization we consider weak magnetic fields, for which the electron Zeeman splitting is much smaller than the thermal energy. Our approach does not require the resonant excitation of specific states, e.g. as in spin orientation protocols with $\Lambda$-scheme~\cite{Kastler1952}.  The dynamic electron spin polarization is based on the violation of the detailed balance (the equality of the rates of the direct and reverse processes) between spin flips in nonequilibrium conditions in weak magnetic fields. We call this effect ``dynamic electron spin polarization'', in similarity with the dynamic nuclear spin polarization gained in nonequilibrium conditions in weak magnetic fields~\cite{abragam,maletinsky2009breakdown}.

The proposed concept is applicable to various systems. Here, we theoretically describe and experimentally demonstrate the dynamic spin polarization of electrons in excitons. It is most pronounced for excitons with the long lifetimes and small splittings between bright and dark states. These conditions are valid for the excitons that are indirect either in real or in momentum space. As a test bed we use momentum indirect excitons confined in (In,Al)As/AlAs QDs, where we get an electron spin polarization of 30\% in a magnetic field of 17~mT.

\section{Microscopic mechanism}
\begin{figure}
  \includegraphics[width=\linewidth]{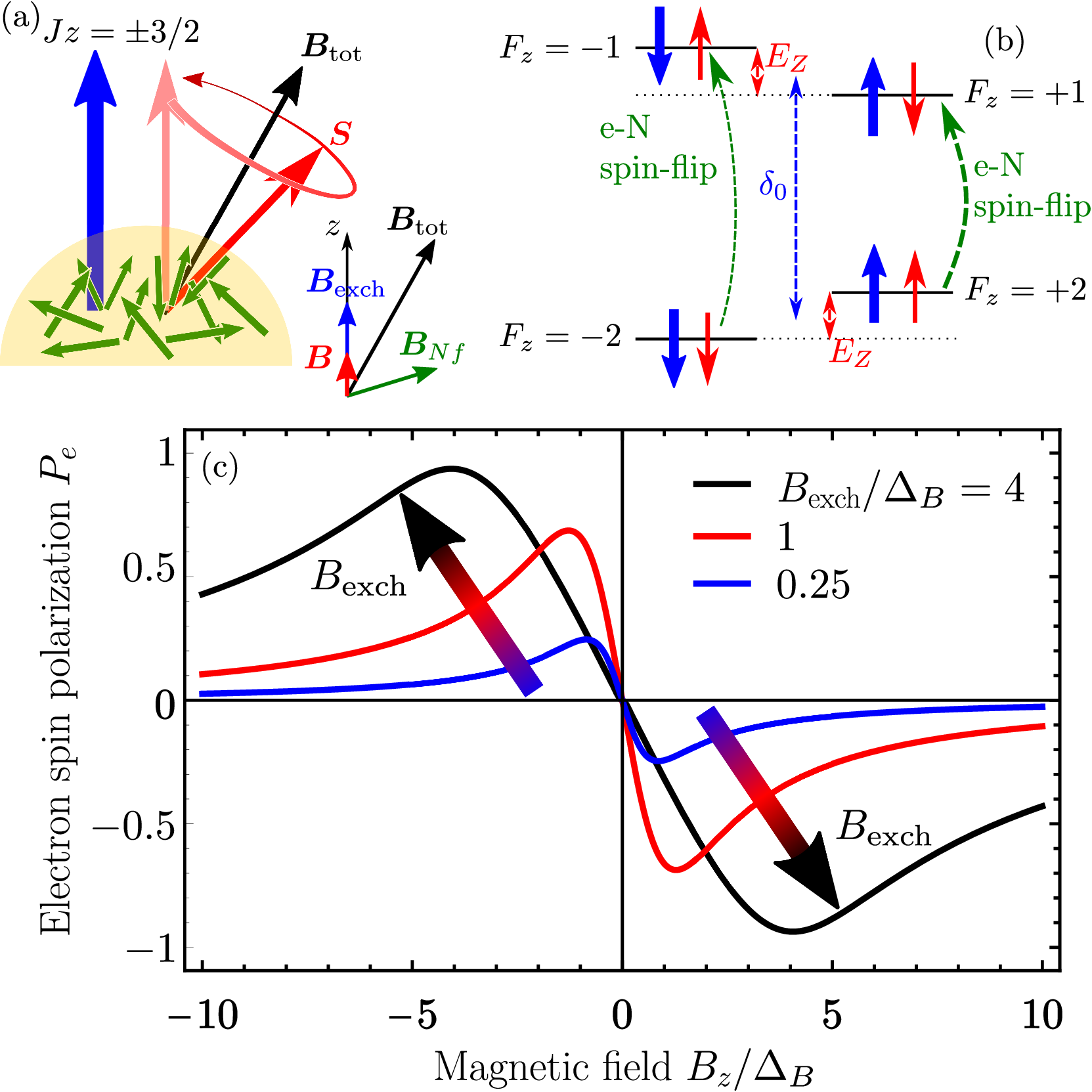}
  \caption{(a) Sketch of the QD (yellow) with randomly oriented nuclear spins (green arrows) and the spins of electron (red) and hole (blue) in a localized exciton. The electron spin precesses in the total field $\bm B_{\rm tot}$.
(b) Fine structure of the exciton levels with electron and heavy hole spins denoted by red and blue arrows, respectively (for $g_h=0$ and $B_{\rm exch}>B_z\gg\Delta_B$). The splitting of the bright ($F_z=\pm1$) and dark states ($F_z=\pm2$) due to the exchange interaction is changed in the external magnetic field by $E_Z=g_e\mu_BB_z$. As a result, the rates of nuclei-assisted spin flips in these pairs of states (green arrows) are different. (c) Electron spin polarization degree as a function of the external magnetic field for the different exchange interaction strengths indicated in the legend, calculated after Eq.~\eqref{eqs:tau_d} with the parameters $g_e\mu_B\Delta_B\tau_b/\hbar=10^3$ and $\tau_s^e/\tau_b=10^3$ (see Ref.~\cite{supp}).}
  \label{fig:sketch}
\end{figure}

We consider an exciton localized in a QD, which consists of an electron with spin projection on the growth $z$-axis $S_z=\pm1/2$, and the heavy hole with spin $J_z=\pm3/2$~\cite{toulouse1}, Fig.~\ref{fig:sketch}(a). We neglect the possible valley degeneracy of the states as well as the interaction between the excitons. The exciton Hamiltonian in the external longitudinal magnetic field $\bm B=(0,0,B_z)$ (Faraday geometry) has the form
\begin{equation}
  \label{eq:H}
  \mathcal H=g_e\mu_B(\bm B+\bm B_{Nf})\bm S+g_h\mu_BB_zJ_z-\frac{2}{3}\delta_0S_zJ_z.
\end{equation}
Here $g_e$ and $g_h$ are the electron and hole longitudinal $g$ factors, respectively, $\mu_B$ is the Bohr magneton, $\bm B_{Nf}$ is the random Overhauser field, $\bm S$ is the spin of electron in exciton. The exchange interaction splits the four exciton states by $\delta_0$ into two upper bright states with the total spin $F_z=S_z+J_z=\pm1$ and the lower dark ones with $F_z=\pm2$. The exciton fine structure is sketched in Fig.~\ref{fig:sketch}(b). The bright excitons can radiatively recombine and have the lifetime $\tau_b$, while the radiative recombination of the dark excitons is spin forbidden. We account for the short-range exchange interaction only and neglect the long-range one. We also neglect the hole hyperfine interaction, which is small due to the $p$-type of the Bloch wave functions~\cite{book_Glazov}. The key difference of our model with the standard description of the exciton states~\cite{toulouse1,noise-excitons} is the hyperfine interaction of electrons in excitons with nuclei, which can be comparable with the exciton exchange splitting $\delta_0$.

The typical time scale of the nuclear spin dynamics is milliseconds, so $\bm B_{Nf}$ can be considered as ``frozen'' for short times~\cite{book_Glazov,merkulov02,PhysRevB.98.121304}. This makes the electron spin relaxation non-Markovian (due to the long nuclear spin memory time), which, as we demonstrate below, leads to the dynamic electron spin polarization. The electron spin precesses in the total magnetic field $\bm B_{\rm tot}=\bm B+\bm B_{Nf}+\bm B_{\rm exch}$, see Fig.~\ref{fig:sketch}(a). Here the exchange field $\bm B_{\rm exch}$ is directed along the heavy hole spin quantization axis $z$: $B_{{\rm exch},z}=-(2/3)\delta_0J_z/(g_e\mu_B)$~\cite{Astakhov07}. The electron spin precession can be described as a classical precession of the magnetic moment, but the exchange field $\bm B_{\rm exch}$ is essentially quantum. It can not be described in the mean field approach, but must be treated as a quantum operator with the two eigenvalues with the opposite signs [see, e.g. Eqs.~\eqref{eq:kinetic_av} below].

The exciton spin dynamics with account for the  incoherent processes can be described in the density matrix formalism. For simplicity, we consider only two such processes: exciton nonresonant generation with the rate $G$ and bright exciton recombination with the time $\tau_b$. The more elaborate model of the spin dynamics is described in Ref.~\onlinecite{supp}. Moreover, we assume that the electron spin precesses around $\bm B_{\rm tot}$ much faster than the bright exciton recombines, so the average electron spin is parallel to $\bm B_{\rm tot}$. We denote the total occupancies of the states with $F_z=+2,+1$ ($-2,-1$) as $N^+$ ($N^-$), and introduce the average electron spins $S^\pm_\parallel$ in the corresponding Hilbert spaces. The kinetic equations for these quantities read~\cite{supp}
\begin{subequations}
  \label{eq:kinetic_av}
  \begin{equation}
    \frac{\d N^\pm}{\d t}=\frac{G}{2}-\frac{N^\pm}{2\tau_b}\pm\frac{S_\parallel^\pm\cos(\theta_\pm)}{\tau_b},
  \end{equation}
  \begin{equation}
    \frac{\d S^\pm_\parallel}{\d t}=-\frac{S^\pm_\parallel}{2\tau_b}\pm\frac{N^\pm\cos(\theta_\pm)}{4\tau_b},
  \end{equation}
\end{subequations}
where $\theta_\pm$ are the angles between $\bm B_{\rm tot}$ and the $z$-axis in the corresponding Hilbert spaces.

In the steady state we solve Eqs.~\eqref{eq:kinetic_av} and obtain the average electron spin along the $z$ axis $S_z=\sum_\pm S_\parallel^\pm\cos(\theta_\pm)$. For a QD ensemble the spin polarization has to be averaged over the Gaussian distribution function of the random Overhauser field ${\propto\exp\left(-B_{Nf}^2/\Delta_B^2\right)}$ with $\Delta_B$ describing the dispersion~\cite{merkulov02}. $\Delta_B$ depends only on the hyperfine interaction constant and QD volume. After averaging we obtain the simple expression for the degree of the dynamic electron spin polarization (see Eq.~\eqref{eq:P} in Ref.~\onlinecite{supp})
\begin{equation}
  \label{eq:simple}
  P_e=\frac{-2B_zB_{\rm exch}}{B_{\rm exch}^2+\Delta_B^2/2+B_z^2}.
\end{equation}
This is the main theoretical result of this Letter.

In Fig.~\ref{fig:sketch}(c) we show the electron spin polarization as  a function of $B_z$ calculated in the extended model accounting for the finite bright and dark exciton lifetimes~\cite{supp}. Noteworthy, it agrees with the simple Eq.~\eqref{eq:simple} within 25\% accuracy. Generally, the electron spin polarization is an odd function of $B_z$ in agreement with the time reversal symmetry. The  polarization reaches maximum at $B_z\approx\sqrt{\Delta_B^2/2+B_{\rm exch}^2}$ and vanishes in large magnetic fields. This is in stark contrast with the thermal spin polarization, which monotonically saturates in strong fields. We note, that the fact, that $P_e$ stays finite in the limit of small random Overhauser field, $\Delta_B\to0$ is related with the above mentioned assumption, that the spin precession frequency in $\bm B_{\rm tot}$ is faster, than the bright exciton recombination. If the exchange splitting  exceeds the typical hyperfine interaction energy (or equivalently $B_{\rm exch}\gtrsim\Delta_B$), then the electron spin polarization can approach 100\%.

Let us qualitatively describe the origin of the dynamic electron spin polarization. The bright excitons recombine during the characteristic time $\tau_b$, while the dark excitons can recombine only due to the nuclei-assisted mixing with the bright states. The transverse components of the random Overhauser field $B_{Nf,x}$ and $B_{Nf,y}$ lead to the electron-nuclear spin flips. The larger the energy difference, between bright and dark states, the smaller the mixing. Note, that this is in contrast with the phonon assisted spin relaxation and is a direct consequence of the non-Markovian spin relaxation. Thus, one can see from Fig.~\ref{fig:sketch}(b), that the mixing is different for the dark states with $F_z=-2$ and $+2$, so one of them recombines faster. Noteworthy, this requires both: exchange splitting between bright and dark states and the longitudinal magnetic field~\cite{supp}. The difference in the lifetimes of the exciton states with spin-up and spin-down electron results in the dynamic polarization of electron spin.

If the bright exciton lifetime is shorter than the typical spin precession period in the random Overhauser field, $\tau_bg_e\mu_B\Delta_B/\hbar\ll1$, then the bright exciton states have large homogeneous broadening, and this leads to the suppression of dynamic electron spin polarization. In the opposite limit, we can use perturbation theory to calculate the lifetimes of the dark excitons with $F_z=\pm2$:
\begin{equation}
  \label{eq:tau_pm}
  \frac{1}{\tau_{\pm2}}=\frac{1}{\tau_b}\frac{B_{Nf,x}^2+B_{Nf,y}^2}{(B_z+B_{Nf,z}\mp B_{\rm exch})^2} .
\end{equation}
The smaller $B_{Nf,x}$ and $B_{Nf,y}$, the longer $\tau_{\pm2}$, so they can be much longer than $\tau_b$. In the steady state the occupancies of the dark states are $N_{\pm2}=(G/4)\tau_{\pm2}$, which are on average much larger than the occupancies of the bight states. As a result, the polarization degree of electron spins  is $(N_{+2}-N_{-2})/(N_{+2}+N_{-2})$, which yields Eq.~\eqref{eq:simple}. The spin polarization degrees of electrons, holes and excitons in this case coincide.

From the above derivation it follows that for pulsed excitation, the dynamic polarization will arise not immediately, but with a delay, e.g., only after the recombination of the bright excitons. This is in stark contrast with the usual optical orientation.

To summarize the theory predictions, the dynamic electron spin polarization for excitons requires: (i) The exciton lifetime to be longer, than the typical electron spin precession period in the random Overhauser field, $\tau_bg_e\mu_B\Delta_B/\hbar>1$. (ii) The exchange interaction between electron and hole to be smaller than the thermal energy. Otherwise, the dynamic electron spin polarization mechanism will smoothly transform into the thermal spin polarization~\cite{supp,Ivchenko2018}. Both requirements can be met using separation between electron and hole either in real or in momentum space.

\begin{figure}
\includegraphics [clip,width=.8\columnwidth]{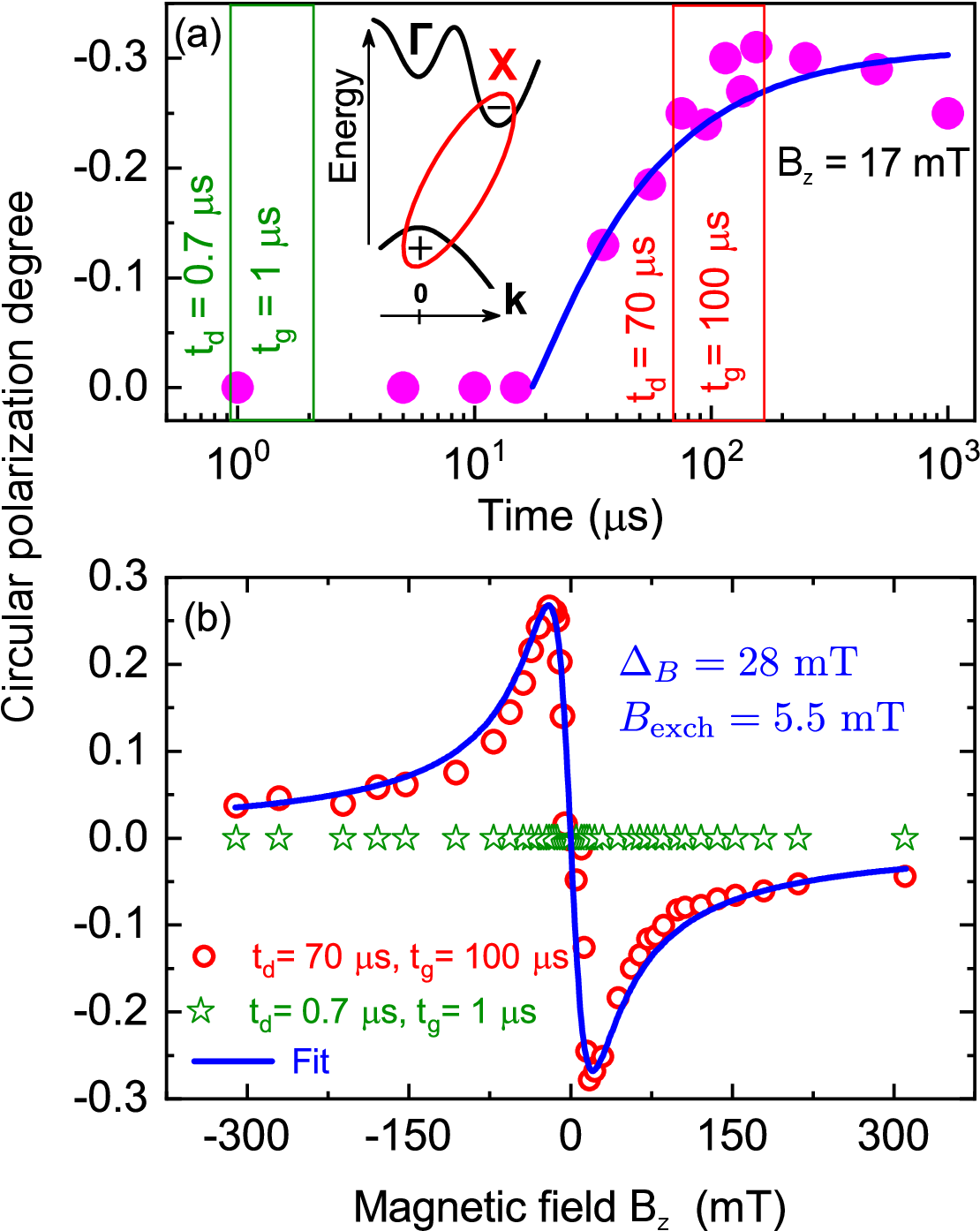}
\caption{(a) Dynamics of PL circular polarization  degree measured at $B_z=17$~mT and $T=2$~K, the integration time is $5~\mu$s. Vertical lines show time-integration windows for panel (b). Blue line is a fit after Eqs.~\eqref{eq:kinetic_av} with parameters $\Delta_B=28$~mT, $B_{\rm exch}=6.6$~mT and $\tau_b=2$~$\mu$s. The inset shows the band diagram of the momentum-indirect (In,Al)As/AlAs QDs. (b) Magnetic field dependencies of the polarization degree measured at: $0.7~\mu$s (green stars) and $70~\mu$s (red circles) with the integration windows of $1~\mu$s and $100~\mu$s, respectively. Blue line is a fit after Eq.~\eqref{eq:simple} with $\Delta_B=28$~mT and $B_{\rm exch}=5.5$~mT.}
\label{fig3}
\end{figure}

\section{Experiment}For experimental demonstration of the suggested mechanism we choose the momentum-indirect (In,Al)As/AlAs QDs. Recently, we showed that in these QDs at low temperatures the exciton spin relaxation is dominated by the hyperfine interaction with $\Delta_B$ being a few millitesla~\cite{PhysRevB.101.075412}, while the exciton lifetime reaches hundreds of microseconds~\cite{Shamirzaev84,doi:10.1063/1.4754619}. The QDs have type-I band alignment (both electron and hole are localized inside the QD)~\cite{ShamirzaevAPL92,Shamirzaev78}. In large QDs the lowest electron and hole states are in the $\Gamma$ valley, so the excitons are momentum-direct~\cite{supp}, but with the decrease of the QD size the $\Gamma$-valley of the conduction band shifts to higher energies faster than the $X$-valley, due to the smaller effective mass and the strain~\cite{Shamirzaev78}. As a result, the electron ground state in small QDs is in the $X$ valley, see inset in Fig.~\ref{fig3}(a), so the excitons in these QDs are momentum-indirect. These excitons, nevertheless, have finite radiative lifetime due to their mixing with the direct excitons at QD interfaces. The spectral distribution of exciton lifetimes allows us to identify the indirect QDs in the inhomogeneous ensemble~\cite{supp}. For photoluminescence (PL) studies we used nonresonant pulsed optical excitation at $3.49$~eV by linearly polarized light. In absence of the magnetic field the exciton PL is unpolarized.

To dynamically polarize electron spins, we apply  the longitudinal magnetic field of $17$~mT (Faraday geometry). In Fig.~\ref{fig3}(a) we show the PL circular polarization as a function of time, detected at the energy of $1.70$~eV (see Ref.~\cite{supp} for the details). The degree of circular polarization is defined as $P_c=(I_+-I_-)/(I_++I_-)$ where $I_\pm$ are the intensities of $\sigma^\pm$ polarized emission. The polarization appears with a delay of $15~\mu$s after the pump pulse and saturates after $100~\mu$s. It is in line with model prediction, that the dynamic polarization appears only after recombination of the bright excitons. Noteworthy, the PL stays polarized up to 1~ms.

The magnetic field dependence of the dynamic polarization integrated for two time windows is shown in Fig.~\ref{fig3}(b). This is the main experimental result of this Letter. The absolute value of $P_{c}(B_z)$ increases in weak fields, reaches maximum of about $0.3$ at $B_z=17$~mT, and then monotonously decreases tending to zero in high fields.  Fitting this dependence with  Eq.~\eqref{eq:simple}  we find  two parameters: $\Delta_B=28$~mT and $B_{\rm exch}=5.5$~mT. The strength of the hyperfine interaction is in good agreement with measurements of optical spin orientation in transverse and longitudinal magnetic fields in a similar sample~\cite{Rautert99,PhysRevB.101.075412}, which supports our interpretation. Using the electron $g$ factor $g_e=2$~\cite{PhysRevB.90.125431,PhysRevB.97.245306}, we find the splitting between bright and dark states $\delta_0=0.6~\mu$eV. This is an unusually small value, which we verified by the PL dynamics in weak transverse magnetic fields~\cite{supp}.

\begin{figure}
\includegraphics [clip,width=0.96\columnwidth]{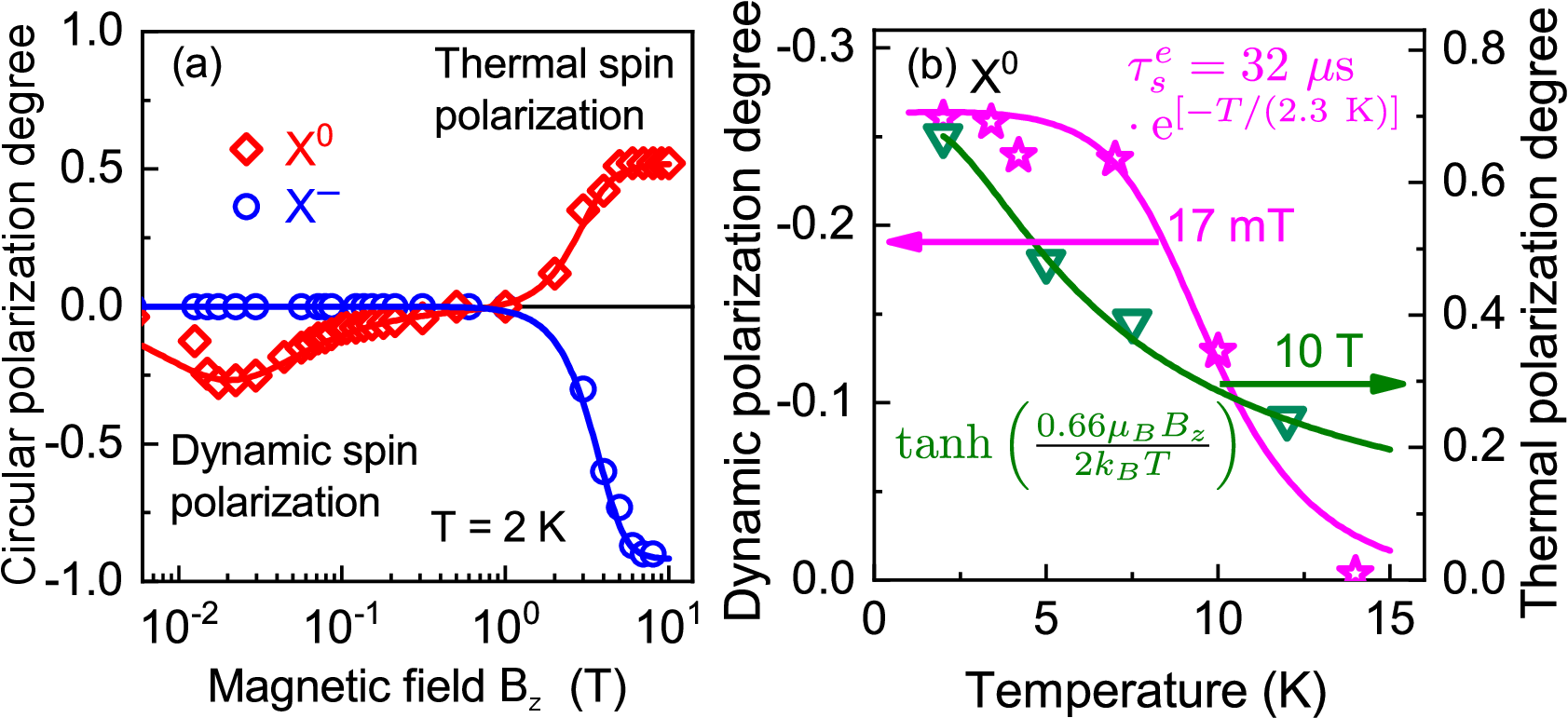}
\caption{(a) PL circular polarization degree as a function of the longitudinal magnetic field for excitons (X$^{0}$) and negatively charged trions (X$^{-}$). (b) Temperature dependence of the dynamic polarization of excitons measured in time window $70-170~\mu$s at $B_z=17$~mT (magenta stars) and the thermal polarization at $B_z=10$~T (green triangles). The fit details for both panels are described in Ref.~\onlinecite{supp}. In particular, the magenta curve in panel (b) is calculated using the same parameters as for the fit in Fig.~\ref{fig3}(a) and an activation law for the electron spin relaxation time given in the panel.} 
\label{fig4}
\end{figure}

To evidence the role of the thermal spin polarization, we measured the polarized PL in magnetic fields up to $10$~T. The results are shown by the red diamonds in Fig.~\ref{fig4}(a). One can see, that the dynamic electron spin polarization takes place in fields $\sim10$~mT only, while the thermal polarization appears in  fields larger than $1$~T and has the opposite (positive) sign.

The blue circles in Fig.~\ref{fig4}(a) show for comparison the PL polarization for an ensemble of negatively charged (In,Al)As/AlAs QDs, see details in Ref.~\cite{supp}. Photoexcitation of singly-charged QDs generates negatively charged excitons (trions)~\cite{doi:10.1063/1.4754619}. In its ground singlet state the electron-hole exchange interaction is absent and, therefore, dynamic electron spin polarization in weak fields does not form, in agreement with the theory. In the same time, the thermal spin polarization has the negative sign~\cite{doi:10.1063/1.4754619}, opposite to the exciton thermal polarization. Note, that the excitation of the triplet trion state would allow one to observe the dynamic electron spin polarization for trions and to transfer it to the resident charge carriers, as we show in Ref.~\onlinecite{supp}.

Additionally, dynamic and thermal spin polarization differ by their temporal dynamics~\cite{supp} and temperature dependencies. The dynamic polarization is temperature independent, as long as the electron spin relaxation is dominated by the hyperfine interaction. By contrast, the thermal polarization is controlled by the factor $E_Z /k_B T$ and  decreases with raising temperature. This is experimentally proven in Fig.~\ref{fig4}(b), where the thermal polarization at $10$~T (green line) decays rapidly with temperature increase, while the dynamic polarization is constant in the range $2-7$~K. We assume that with further temperature increase the phonon-assisted electron spin relaxation time $\tau_s^e$ shortens, so the dynamic polarization decays at the time scales longer than $\tau_s^e$. As a result, when $\tau_s^e$ becomes shorter than the delay after the pump pulse, the spin polarization decreases.

Realization of the dynamic electron spin polarization in typical direct GaAs QDs is prevented by the large splitting between bright and dark excitons and by short exciton lifetimes. These limitations can be overcome in type-II QDs or core-shell colloidal nanocrystals~\cite{raino2011probing}. Another promising platform for the implementation of the dynamic spin polarization are twisted heterobilayers of transition metal dichalcogenides. The moir\'e pattern in these structures creates a superlattice potential with the typical period of about $5$~nm~\cite{seyler2019signatures}. For excitons localized in this potential, the hyperfine interaction leads to the spin-valley relaxation time of the order of tens of nanoseconds~\cite{MX2_Avdeev}, while the exciton lifetime due to the confinement and spatial separation of electron and hole can be as long as~$100$~ns~\cite{rivera2016valley,seyler2019signatures}. The fine stricture of spin singlet and spin triplet excitons confined in a moir\'e potential remains poorly investigated~\cite{Yu_2018,forg2019cavity}. Nevertheless, due to the electron-hole separation in heterobilayers the fine structure splitting is expected to be small, and can be of the order of the hyperfine interaction strength. Therefore we expect the dynamic spin polarization of moir\'e trapped excitons.

In the process of the dynamic electron spin polarization, the angular momentum for the electrons is gained from the nuclear spin bath via the hyperfine interaction. This looks surprising, as commonly this interaction is considered as a source of spin relaxation only~\cite{merkulov02,Bechtold2015}. The dynamic electron polarization can be further transferred to nuclei~\cite{korenev1999dynamic} or  magnetic impurities.
Additionally, the dynamic spin polarization can be transferred to the resident electrons in the charged QDs~\cite{supp}. The dynamic electron spin polarization has the following advantages: (i) It requires weak magnetic fields, which can be easily modulated. (ii) It uses nonresonant and unpolarized optical excitation. (iii) It is temperature independent, as long as the spin relaxation is dominated by the hyperfine interaction. (iv) 100\% electron spin polarization is feasible. All that makes dynamic spin polarization very attractive for the spin orientation in nanodevices for quantum information processing.

We thank E. L. Ivchenko with M. M. Glazov for fruitful
discussions, J.~Rautert with D.~Kudlacik for technical support, RF
President Grant No. MK-1576.2019.2 and the Basis Foundation.
This work was supported by the Deutsche Forschungsgemeinschaft via
the project No. 409810106 and by the Russian Foundation for Basic
Research Grant No.19-52-12001. T.S.Sh. acknowledges the financial
support by the Russian Foundation for Basic Research (Grants No.
19-02-00098), and by the Act 211 Government of the Russian
Federation (Contract No.02.A03.21.0006). Government of the Russian
Federation also supports this work via Grant No.
AAAA-A17-117042110141-5. The theoretical studies were supported by the Russian Science Foundation
(Grant No. 19-72-00081). M.B. acknowledges support by the Deutsche Forschungsgemeinschaft (TRR 160, project A1).

\renewcommand{\i}{\ifr}
\let\oldaddcontentsline\addcontentsline
\renewcommand{\addcontentsline}[3]{}

\let\addcontentsline\oldaddcontentsline
\makeatletter
\renewcommand\tableofcontents{%
    \@starttoc{toc}%
}
\makeatother
\renewcommand{\i}{{\rm i}}

\onecolumngrid
\vspace{\columnsep}
\begin{center}
\newpage
\makeatletter
{\large\bf{Supplemental Material to\\``\@title''}}
\makeatother
\end{center}
\vspace{\columnsep}

The supplementary information presents the following topics:

\hypersetup{linktoc=page}
\tableofcontents
\vspace{\columnsep}
\twocolumngrid

\renewcommand{\section}[1]{\oldsec{#1}}
\renewcommand{\thepage}{S\arabic{page}}
\renewcommand{\theequation}{S\arabic{equation}}
\renewcommand{\thefigure}{S\arabic{figure}}
\renewcommand{\bibnumfmt}[1]{[S#1]}
\renewcommand{\citenumfont}[1]{S#1}

\setcounter{page}{1}
\setcounter{section}{0}
\setcounter{equation}{0}
\setcounter{figure}{0}

\section{S1. Extended theoretical model}

We consider the four states of an exciton localized in a QD with the electron spin $S_z=\pm1/2$ and heavy hole spin $J_z=\pm3/2$ ($z$ is the QD growth axis). The total angular momentum is $F_z=S_z+J_z=\pm1$ for the bright exciton states and $F_z=\pm2$ for the dark exciton states. The exciton Hamiltonian has the form [Eq.~\eqref{eq:H} in the main text]
\begin{equation}
  \mathcal H=g_e\mu_B(\bm B+\bm B_{Nf})\bm S+g_h\mu_BB_zJ_z-\frac{2}{3}\delta_0s_zJ_z.
\end{equation}
Here $g_e$ and $g_h$ are the electron and hole longitudinal $g$ factors, respectively, $\mu_B$ is the Bohr magneton, $\bm B_{Nf}$ is the random Overhauser field (assumed to be static), $\bm S$ is the electron spin, and $\delta_0$ is the electron-hole exchange interaction energy.

We describe the exciton spin dynamics using the density matrix formalism. The density matrix $\rho(t)$ satisfies the master equation
\begin{equation}
  \frac{\d\rho(t)}{\d t}=-\frac{\i}{\hbar}\left[\mathcal H,\rho\right]-\mathcal L\left\{\rho\right\},
\end{equation}
where $\mathcal L$ is a linear superoperator describing the incoherent processes. These are the exciton generation with the rate $G$, spin relaxation of electrons and holes in excitons with the times $\tau_s^e$ and $\tau_s^h$, respectively, spin flips between bright (dark) states with the time $\tau_1$ ($\tau_2$), radiative recombination of the bright excitons with the time $\tau_R$, and nonradiative recombination of the dark excitons with the time $\tau_{NR}$. The density matrix is characterized by the numbers of the bright and dark excitons, $N_b$ and $N_d$, respectively, by the average spin of the electron in exciton, $\bm S$, by the heavy hole spin component, $J_z$, and by the electron-hole spin correlation
\begin{equation}
  \label{eq:Q}
  \bm Q=\frac{2}{3}\bm SJ_z.
\end{equation}
These are the time-dependent average values, which
can be calculated as the traces of the corresponding operators
multiplied by the density matrix $\rho(t)$. They satisfy the
following set of kinetic equations~\cite{S_noise-excitons}:
\begin{widetext}
\begin{subequations}
  \label{eqs:complete}
  \begin{equation}
  \frac{\d N_b}{\d t}=\frac{G}{2}-\frac{N_b}{\tau_R}+Q_z\left(\frac{1}{\tau_s^h}+\frac{1}{\tau_s^e}\right)-Q_y\Omega_x+Q_x\Omega_y,
  \end{equation}
  \begin{equation}
  \frac{\d N_d}{\d t}=\frac{G}{2}-\frac{N_d}{\tau_{NR}}-Q_z\left(\frac{1}{\tau_s^h}+\frac{1}{\tau_s^e}\right)+Q_y\Omega_x-Q_x\Omega_y,
  \end{equation}
  \begin{equation}
  \frac{\d S_x}{\d t}=-S_x\left(\frac{1}{2\tau_b}+\frac{1}{2\tau_d}+\frac{1}{\tau_s^e}\right)-S_y\Omega_z+S_z\Omega_y+Q_y\frac{\delta_0}{\hbar},
  \end{equation}
 \begin{equation}
  \frac{\d S_y}{\d t}=-S_y\left(\frac{1}{2\tau_b}+\frac{1}{2\tau_d}+\frac{1}{\tau_s^e}\right)+S_x\Omega_z-S_z\Omega_x-Q_x\frac{\delta_0}{\hbar},
  \end{equation}
  \begin{equation}
  \frac{\d S_z}{\d t}=-\frac{S_z}{\tau_s^e}-\frac{S_z-J_z/3}{2\tau_b}-\frac{S_z+J_z/3}{2\tau_d}-S_x\Omega_y+S_y\Omega_x,
  \end{equation}
  \begin{equation}
  \frac{\d J_z}{\d t}=-\frac{J_z}{\tau_s^h}-\frac{J_z-3S_z}{2\tau_b}-\frac{J_z+3S_z}{2\tau_d},
  \end{equation}
  \begin{equation}
  \frac{\d Q_x}{\d t}=-Q_x\left(\frac{1}{2\tau_b}+\frac{1}{2\tau_d}+\frac{1}{\tau_s^e}+\frac{1}{\tau_s^h}\right)-Q_y\Omega_z+Q_z\Omega_y+S_y\frac{\delta_0}{\hbar},
  \end{equation}
  \begin{equation}
  \frac{\d Q_y}{\d t}=-Q_y\left(\frac{1}{2\tau_b}+\frac{1}{2\tau_d}+\frac{1}{\tau_s^e}+\frac{1}{\tau_s^h}\right)-Q_z\Omega_x+Q_x\Omega_z-S_x\frac{\delta_0}{\hbar},
  \end{equation}
\end{subequations}
\end{widetext}
where $1/\tau_b=1/\tau_R+1/\tau_1$, $1/\tau_d=1/\tau_{NR}+1/\tau_2$, and $\bm\Omega=g_e\mu_B(\bm B+\bm B_{Nf})/\hbar$. Note that $Q_z=(N_d-N_b)/2$, as follows from Eq.~\eqref{eq:Q}. Note also, that in the main text we use a simplified model and, in particular, neglect the exciton spin flip time $\tau_1$, so that $\tau_b=\tau_R$ in the main text. The solution of these equations allows one to describe the dynamic electron spin polarization, for example, in the steady state. For pulsed excitation these equations should be solved with zero initial conditions except for $N_b=N_d=N/2$ with $N$ being the total number of the created excitons, because the density matrix immediately after the pump pulse is proportional to the identity matrix.

The random Overhauser field, $\bm B_{Nf}$, is different for the different QDs, so the solution of the kinetic equations should be averaged over its distribution function
\begin{equation}
  \label{eq:F}
  \mathcal F(\bm B_{Nf})=\frac{1}{(\sqrt{\pi}\Delta_B)^3}\exp\left(-\frac{B_{Nf}^2}{\Delta_B^2}\right).
\end{equation}
The electron spin polarization degree is given by
\begin{equation}
  \label{eq:P}
  P_e=\frac{2\braket{S_z}}{\braket{N_b+N_d}},
\end{equation}
where the angular brackets denote the averaging with the distribution function~\eqref{eq:F}. The intensities of the emitted circularly polarized light $I_\pm$ are proportional to the number of excitons recombining per unit time, and they should be averaged as well:
\begin{equation}
  I_\pm\propto\braket{\frac{N_b}{2}\pm\frac{1}{2}\left(\frac{J_z}{3}-S_z\right)}\frac{1}{\tau_R}.
\end{equation}

The kinetic equations are greatly simplified if one neglects the exciton spin flips ($\tau_1$ and $\tau_2$) and the hole spin relaxation ($\tau_s^h$). In this limit we introduce $N^\pm=(N_b+N_d)/2\mp J_z/3$ and $\bm S^\pm=(\bm S\mp\bm Q)/2$, and obtain from Eqs.~\eqref{eqs:complete} the following kinetic equations:
\begin{subequations}
  \label{eqs:tau_d}
  \begin{equation}
    \frac{\d N^\pm}{\d t}=\frac{G}{2}-\frac{N^\pm}{2}\left(\frac{1}{\tau_b}+\frac{1}{\tau_d}\right)\pm S_z^\pm\left(\frac{1}{\tau_b}-\frac{1}{\tau_d}\right),
  \end{equation}
  \begin{multline}
    \frac{\d\bm S^\pm}{\d t}=\bm\Omega^\pm \times\bm S^\pm-\frac{\bm S^\pm}{2}\left(\frac{1}{\tau_b}+\frac{1}{\tau_d}+\frac{2}{\tau_s^e}\right)\\\pm\frac{N^\pm}{4}\left(\frac{1}{\tau_b}-\frac{1}{\tau_d}\right)\bm e_z.
  \end{multline}
\end{subequations}
Here $\bm\Omega^\pm=g_e\mu_B\bm B_{\rm tot}/\hbar$ are the frequencies of the spin precession of electron in exciton with the heavy-hole spin $J_z=\pm3/2$ and $\bm B_{\rm tot}=\bm B+\bm B_{Nf}-(2/3)\delta_0J_z/(g_e\mu_B)\bm e_z$ with $\bm e_z$ being the unit vector along $z$ axis.

\begin{figure}
  \centering
  \includegraphics[width=\linewidth]{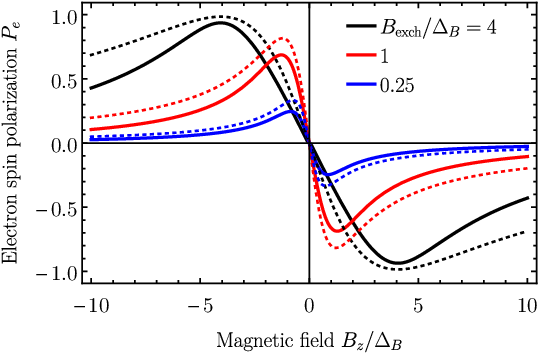}
\caption{Dynamic electron spin polarization as function of magnetic
field. Solid curves are calculated after Eqs.~\eqref{eqs:tau_d} with the parameters $g_e\mu_B\Delta_B\tau_b/\hbar=10^3$, $\tau_s^e/\tau_b=10^3$, $\tau_d\to\infty$ and reproduce the curves from Fig.~\ref{fig:sketch}(c) of the main text, the dashed lines are calculated after Eq.~\eqref{eq:simple_S} with the exchange interaction given in the legend.}
  \label{fig:anal}
\end{figure}

Similarly to the main text, let us assume the fast spin precession, $\Omega^\pm\gg1/\tau_{b,d},1/\tau_s^e$. In this case we arrive at
\begin{subequations}
  \label{eq:fast_all}
  \begin{equation}
    \frac{\d N^\pm}{\d t}=\frac{G}{2}-\frac{N^\pm}{2}\left(\frac{1}{\tau_b}+\frac{1}{\tau_d}\right)\pm S_\parallel^\pm\cos(\theta_\pm)\left(\frac{1}{\tau_b}-\frac{1}{\tau_d}\right),
  \end{equation}
  \begin{equation}
    \frac{\d S^\pm_\parallel}{\d t}=-\frac{S^\pm_\parallel}{2}\left(\frac{1}{\tau_b}+\frac{1}{\tau_d}+\frac{2}{\tau_s^e}\right)\pm\frac{N^\pm\cos(\theta_\pm)}{4}\left(\frac{1}{\tau_b}-\frac{1}{\tau_d}\right),
  \end{equation}
\end{subequations}
where $\theta_\pm$ are the angles between $\bm\Omega^\pm$ and $\bm e_z$.

In the limit $\tau_s,\tau_d\gg\tau_b$ we obtain Eqs.~\eqref{eq:kinetic_av} of the main text:
\begin{subequations}
  \label{eq:kinetic_av_S}
  \begin{equation}
    \frac{\d N^\pm}{\d t}=\frac{G}{2}-\frac{N^\pm}{2\tau_b}\pm\frac{S_\parallel^\pm\cos(\theta_\pm)}{\tau_b},
  \end{equation}
  \begin{equation}
    \frac{\d S^\pm_\parallel}{\d t}=-\frac{S^\pm_\parallel}{2\tau_b}\pm\frac{N^\pm\cos(\theta_\pm)}{4\tau_b}.
  \end{equation}
\end{subequations}
In the steady state we obtain the degree of dynamic electron spin polarization in exciton [Eq.~\eqref{eq:simple} in the main text]:
  \begin{equation}
    \label{eq:simple_S}
    P_e=\frac{-2B_zB_{\rm exch}}{B_{\rm exch}^2+\Delta_B^2/2+B_z^2}.
  \end{equation}
In Fig.~\ref{fig:anal} we compare this expression with the exact calculation after Eqs.~\eqref{eqs:tau_d} with the parameters $g_e\mu_B\Delta_B\tau_b/\hbar=10^3$, $\tau_s^e/\tau_b=10^3$ and $\tau_d\to\infty$. The sizable difference between the curves (up to 25\%) indicates, that there are logarithmic corrections to Eq.~\eqref{eq:simple_S}.

\begin{figure}
  \includegraphics[width=\linewidth]{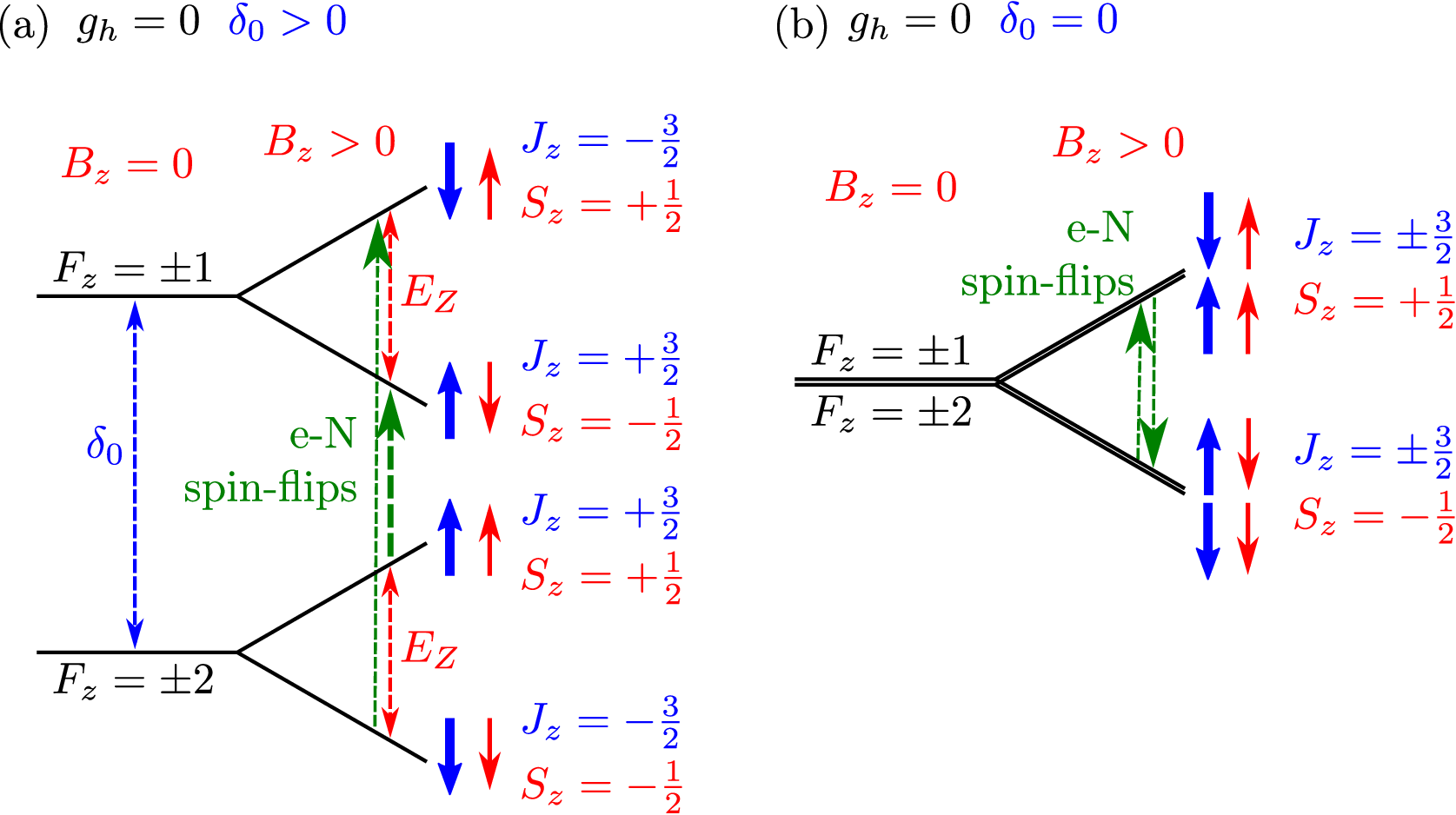}
  \caption{Illustration of the mechanism of dynamic electron spin polarization. (a) Due to the electron-hole exchange interaction $\delta_0>0$ the spin flips shown by the green arrows require the different energies. (b) In the absence of exchange interaction ($\delta_0=0$) the rates of the spin flips are equal.}
  \label{fig:mechanism}
\end{figure}

The mechanism of dynamic electron spin polarization is illustrated in Fig.~\ref{fig:mechanism}. Here we consider for simplicity $g_h=0$ and $\Delta_B\ll B_z$. Panel (a) corresponds to the strong exchange interaction $B_{\rm exch}>B_z$ similarly to Fig.~\ref{fig:sketch}(b) in the main text. After the pump pulse the bright excitons quickly recombine during the time $\sim\tau_b$. The dark exciton recombination requires the electron spin flip, which is shown by the green dashed arrows. The energy difference between the states with $F_z=+2$ and $+1$ is smaller than between the states with $F_z=-2$ and $-1$, so the spin flips for electrons with $S_z=+1/2$ are faster than for $S_z=-1/2$. As a result, the electrons get dynamically polarized.

Figure~\ref{fig:mechanism}(b) illustrates the case without exchange interaction $B_{\rm exch}=0$. In this case the energy difference between the states described above is the same, so the electron spin flips shown by the green arrows have the same rate. This demonstrates, that the electron-hole exchange interaction is necessary to get dynamic electron spin polarization.

Noteworthy, the electron spin flips take place simultaneously with the exciton recombination. This process is possible due to the mixing between electron spin states by the transverse components of the static random Overhauser field. The larger the energy difference between the spin states, the smaller the mixing. This is a direct consequence of the non-Markovian spin relaxation (``frozen'' nuclear field approximation) and in contrast with the phonon-assisted spin flips, which are caused by the quickly fluctuating strain fields acting on the electrons. The spin flip transitions with the absorption or emission of the same energy have the same rates.

\begin{figure}
  \centering
  \includegraphics[width=\linewidth]{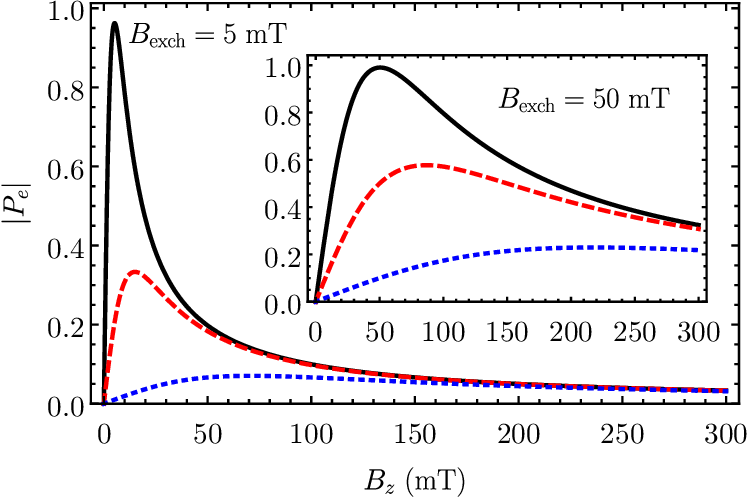}
\caption{Electron spin polarization degree as a function of the external magnetic field calculated after Eq.~\eqref{eq:simple_S} with the parameters $B_{\rm exch}=5$~mT and $\Delta_B=2$, $20$ and~$100$~mT for the solid black, red dashed and blue dotted curves, respectively. The inset shows the calculations for $B_{\rm exch}=50$~mT and $\Delta_B=10$, $100$ and $300$~mT for the solid black, red dashed and blue dotted curves, respectively. }
  \label{fig:anal_exp}
\end{figure}

Let us study the dependence of the dynamic electron spin polarization on the strength of the hyperfine interaction and external magnetic field. In Fig.~\ref{fig:anal_exp} we consider $B_{\rm exch}=5$~mT and show that the polarization is the largest, if the hyperfine interaction is weaker, than the exchange interaction, $\Delta_B<B_{\rm exch}$. In this case the maximum spin polarization is reached at $B_z^{(\rm max)}\sim B_{\rm exch}$. With increase of the strength of the hyperfine interaction, the maximum spin polarization decreases, when $\Delta_B$ becomes larger than $B_{\rm exch}$, the magnetic field dependence becomes smoother and its maximum shifts to $B_z^{(\rm max)}\sim \Delta_B$.  Generally, the maximum spin polarization is reached at
  \begin{equation}
    \label{eq:B_max}
    B_z^{(\rm max)}=\sqrt{B_{\rm exch}^2+\Delta_B^2}.
  \end{equation}
The inset in Fig.~\ref{fig:anal_exp} shows that the same behavior takes place also for the stronger exchange interaction $B_{\rm exch}=50$~mT, but the corresponding hyperfine field fluctuations in this case should be stronger in order to decrease the spin polarization degree. We stress, that for reaching 100\% electron spin polarization the nuclear field fluctuations should be small: $\Delta_B<B_{\rm exch}$, on one hand, but should be the dominant spin relaxation mechanism on the other hand.

\section{S2. Experimental results}

\subsection{A. Experimentals}

The studied self-assembled (In,Al)As QDs embedded in an AlAs matrix
(AG3686)  were grown by the molecular-beam epitaxy on a
semi-insulating $(001)$-oriented GaAs substrate with 400-nm-thick
GaAs buffer layer~\cite{S_Shamirzaev78}. The structure contains one
QD layer sandwiched between two 70-nm-thick AlAs layers. The
nominal amount  of  deposited  InAs is about $2.4$ monolayers. The
top AlAs barrier is protected from the oxidation by the
$20$-nm-thick GaAs cap layer.

\begin{figure}
  \centering
  \includegraphics[width=\linewidth]{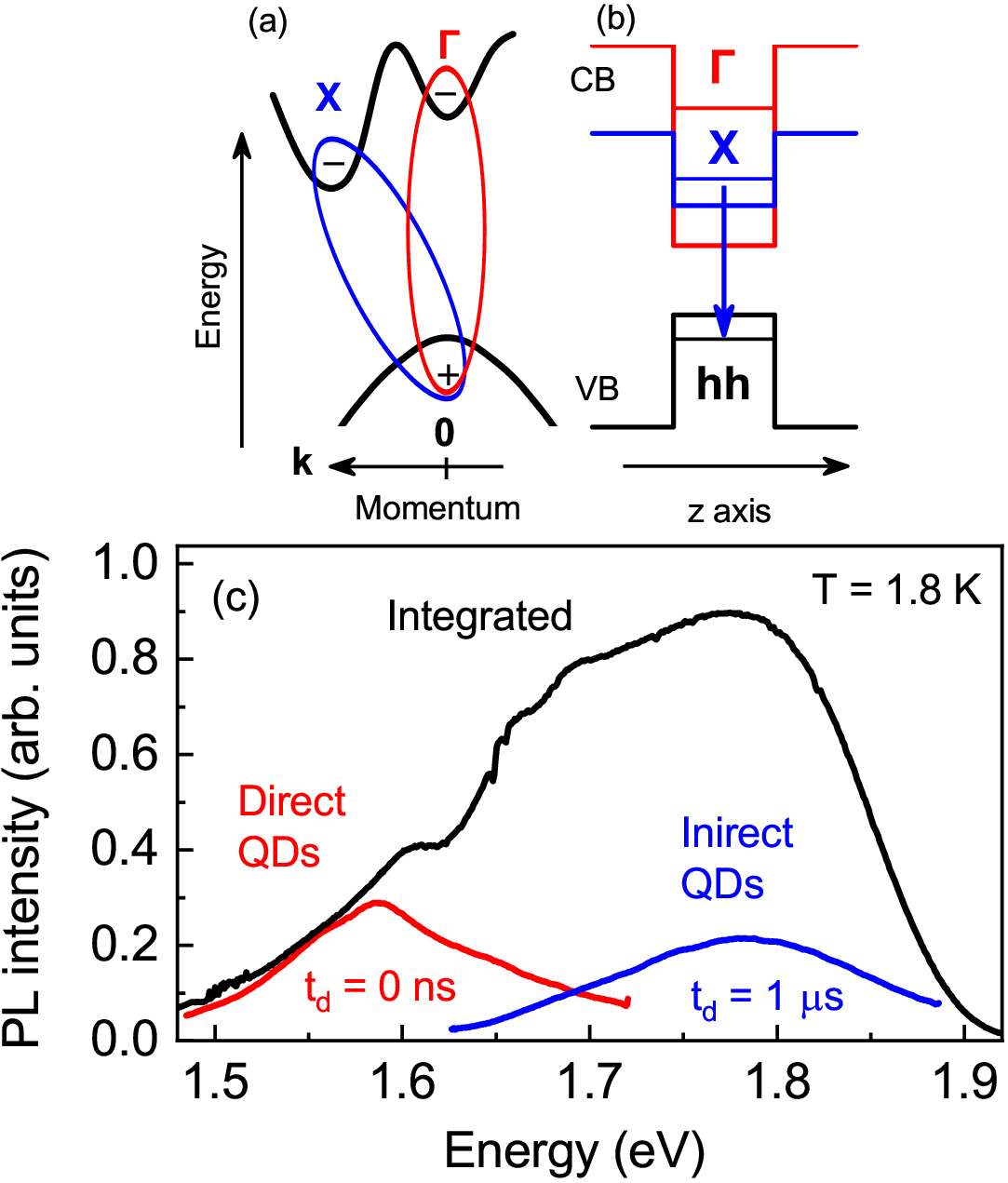}
\caption{Schematic band diagrams of (In,Al)As/AlAs
QDs. (a) Momentum-direct (red) and indirect (blue) excitons consist
of an electron in the $\Gamma$ and X valley, respectively, and a
heavy hole in the $\Gamma$ valley. (b) Type-I band alignment of QDs in real
space. The X valley has lower energy than the $\Gamma$ valley in
small QDs. The arrow shows the ground state optical transition. (c)
PL spectra of (In,Al)As/AlAs QDs measured under nonresonant pulsed
excitation. The black line shows the time-integrated spectrum. The
red and blue lines show the spectra integrated within $t_{\text{g}}=
1$~ns immediately after the pump pulse ($t_{\text{d}}= 0$) and
within $t_{\text{g}} = 200~\mu$s with the delay $t_{\text{d}}=
1~\mu$s after the pump pulse, respectively.}
  \label{fig:PL-spec}
\end{figure}

The QDs can be momentum-direct or momentum-indirect
(see Fig.~\ref{fig:PL-spec}(a)) depending on QD size and have type-I
band alignment (both electron and hole are localized inside the
QD)~\cite{S_Shamirzaev78} as it is shown in Fig.~\ref{fig:PL-spec}(b).

The photoluminescence (PL) was excited nonresonantly with the photon energy of the laser exceeding the direct band gap of the AlAs matrix. We used the third harmonic of a Q-switched Nd:YVO$_4$ pulsed laser with the photon energy $3.49$~eV, pulse duration $5$~ns and the repetition rate 1~kHz~\cite{S_Shamirzaev84}. The excitation was linearly polarized, and we checked that the direction of the polarization does not affect the presented results, as expected for the strongly nonresonant excitation. The PL was dispersed by a $0.5$-m monochromator. For the time-resolved and time-integrated PL measurements we used a gated charge-coupled-device (CCD) camera synchronized with the laser via an external trigger signal. The  time between the pump pulse and the start of the PL recording, $t_{\text{d}}$, could be varied from zero up to $1$~ms. The duration of the PL recording (i.e. the gate window $t_{\text{g}}$) could be varied from $1$~ns to $1000~\mu$s. The signal intensity and the time resolution of the setup depend on $t_{\text{d}}$ and $t_{\text{g}}$. The highest time resolution of the detection system is $1$~ns.

The circular polarization degree of the PL is given by $P_c =(I_{+}-I_{-})/(I_{+}+ I_{-})$, where $I_{+}$ and $I_{-}$ are the intensities of the $\sigma^{+}$ and $\sigma^{-}$ polarized PL components, respectively. To determine the sign of $P_c$, we performed the control measurement on a diluted magnetic semiconductor structure with (Zn,Mn)Se/(Zn,Be)Se quantum wells. For this structure, $P_c>0$ at $B_z>0$ in the Faraday geometry~\cite{S_Keller}.

\subsection{B. Time-resolved photoluminescence}

The time-integrated PL spectrum is shown by the  black line in
Fig.~\ref{fig:PL-spec}(c). In addition, the red line shows the
spectrum immediately after the pump pulse ($t_{\text{d}}=0$~ns)
recorder during $t_{\text{g}}=$1~ns. It demonstrates, that at the
lower energies the exciton recombination is as fast as several
nanoseconds, which corresponds to the momentum-direct QDs. The blue
line shows the spectrum with the delay $t_{\text{d}}=1~\mu$s
recorded during $t_{\text{g}}=$200~$\mu$s. It illustrates the fact
that at higher energies the exciton recombination takes place on a
scale of a few hundreds of $\mu$s, which corresponds to the
momentum-indirect QDs~\cite{S_Shamirzaev84}.

\subsection{C. Polarized photoluminescence}

\begin{figure}[t]
  \centering
  \includegraphics[width=0.6\linewidth]{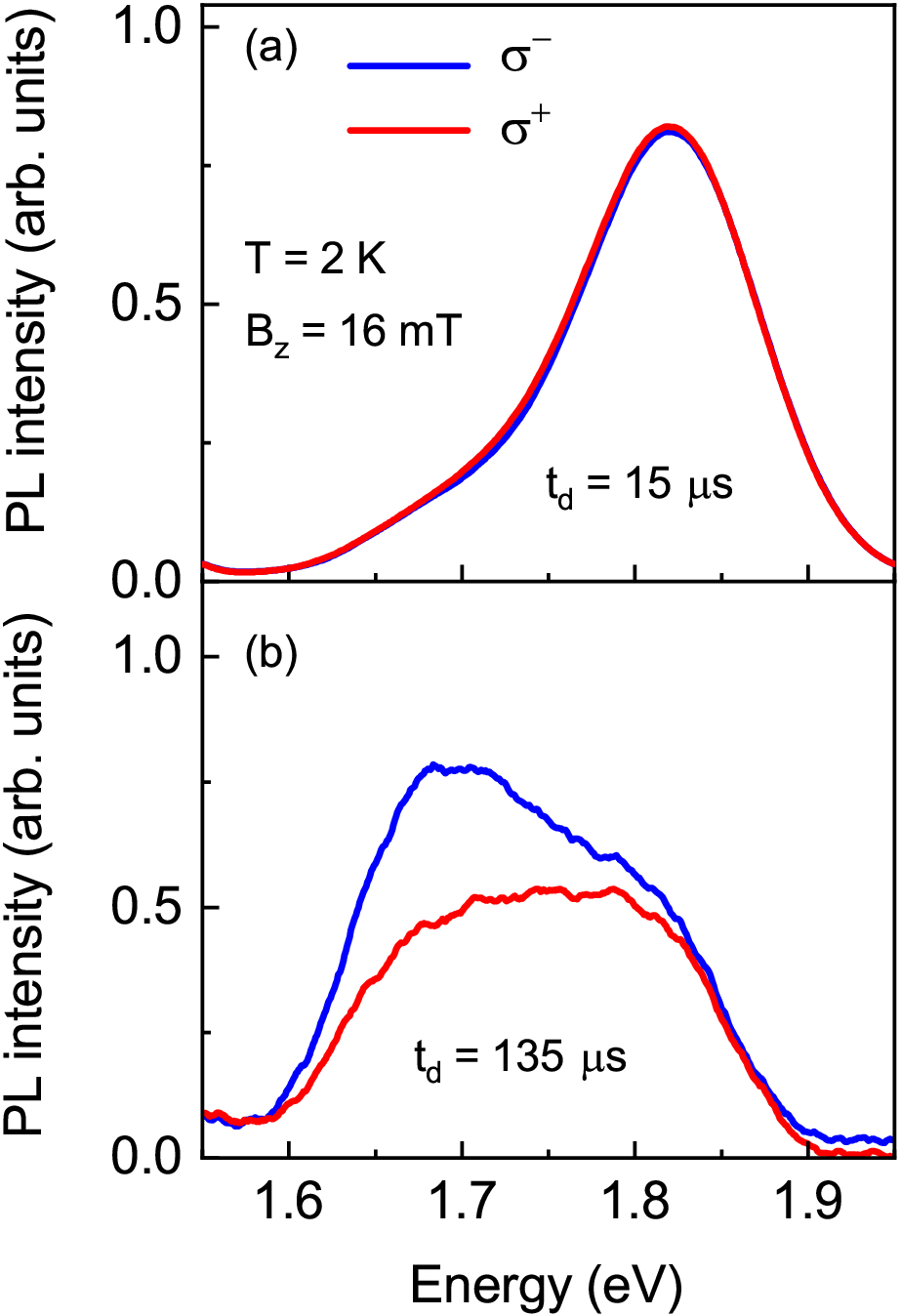}
\caption{PL spectra measured in the opposite circular polarizations (red in $\sigma^+$ and blue in $\sigma^-$ polarization) with the parameters: (a) $t_{\rm d}=15$~$\mu$s and $t_{\rm g}=1$~$\mu$s; (b) $t_{\rm d}=135$~$\mu$s and $t_{\rm g}=100$~$\mu$s.}
  \label{fig:PL_polarized}
\end{figure}

In Figure~\ref{fig:PL_polarized} we show the PL spectra recorded in the opposite circular polarizations at $B_z=16$~mT for the two sets of parameters: $t_{\text{d}}=15~\mu$s, $t_{\text{g}}=1~\mu$s and $t_{\text{d}}=135~\mu$s, $t_{\text{g}}=100~\mu$s. One can see that the PL is unpolarized up to 15~$\mu$s, but at later times it gets negatively polarized.

\begin{figure}[t]
  \centering
  \includegraphics[width=0.8\linewidth]{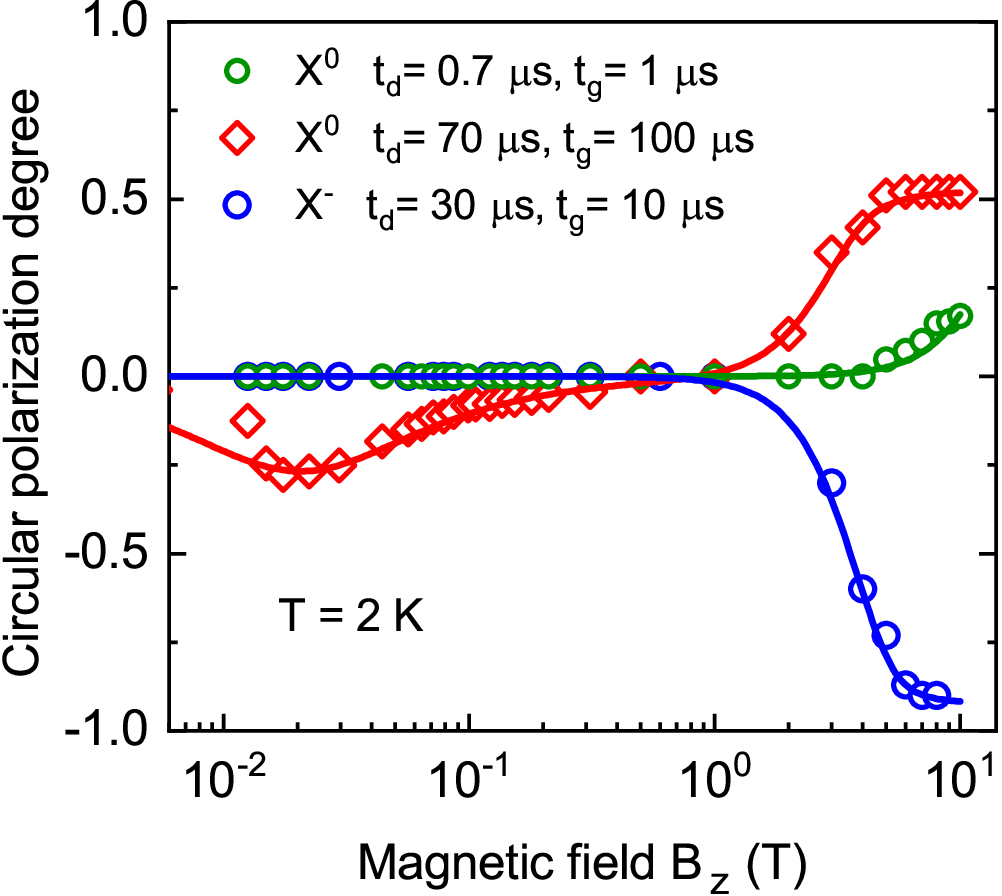}
\caption{Dependencies of $P_c$ on the magnetic field in the large range from zero to $10$~T at short and long time scales for neutral exciton ($X^0$) at $1.67$~eV, as well as at long time scales for negatively charged trion ($X^{-}$) at $1.71$~eV. The fits are described in Sec.~\hyperref[sec:T_fit]{S3.B}.}
  \label{fig:B_dep}
\end{figure}

In Figure~\ref{fig:B_dep} the green circles and red diamonds show the PL polarization degree as a function of the longitudinal magnetic field. It is recorded for the two sets of parameters $t_{\text{d}}=0.7~\mu$s, $t_{\rm g}=1~\mu$s and $t_{\text{d}}=70~\mu$s, $t_{\rm g}=100~\mu$s. For the short delay the circular polarization is absent at low magnetic fields. It appears for the magnetic fields above 4~T only and increases up to $0.17$ at $10$~T. In the same time, for the long delay there is the negative circular polarization at low magnetic fields (up to $0.3$ at 17~mT), which decreases down to 0 with increasing magnetic field up to $1$~T. With further increase of the field, the positive circular polarization appears for the magnetic field above 2~T and $P_c$ saturated at $0.52$ at $B_z\ge5$~T.

For short delays, the PL is related with the recombination of the bright excitons. In this case, according to the theory predictions, there is no electron spin polarization and no circular polarization of the PL in low magnetic fields. The circular polarization in high magnetic fields is induced by the thermal spin polarization.

For long time delays the PL is related with the recombination of the dark excitons through their mixing with the bright excitons. In this case the dynamic electron spin polarization takes place and PL is circularly polarized in small magnetic fields. In the large fields, the differences of the spin polarization in magnitude and in saturation (absent for the short times and present for the long times) show that the spin relaxtion time in high magnetic fields has a value in the range from $1~\mu$s to $70~\mu$s.

\subsection{D. Doped QDs}

The dynamic electron spin polarization requires the exchange interaction between electron and hole. This theoretical prediction can be verified using the QDs with resident electrons. The optical excitation of theses QDs leads to the formation of the singlet trions, where the electron-hole exchange interaction is absent.

We studied the structure with ten layers of (In,Al)As QDs separated by AlAs barriers with the thickness $30$~nm. The QDs were formed during $40$~s at the temperature $515^{\circ}$C. In order to charge them with a single electron on average, a delta-layer of silicon donors with the density $1\times10^{12}$~cm$^{-2}$ was placed in AlAs layers 2~nm below each QD layer. A 15-nm-thick GaAs cap layer protected the top AlAs layer against oxidation.

\begin{figure}
  \centering
  \includegraphics[width=0.6\linewidth]{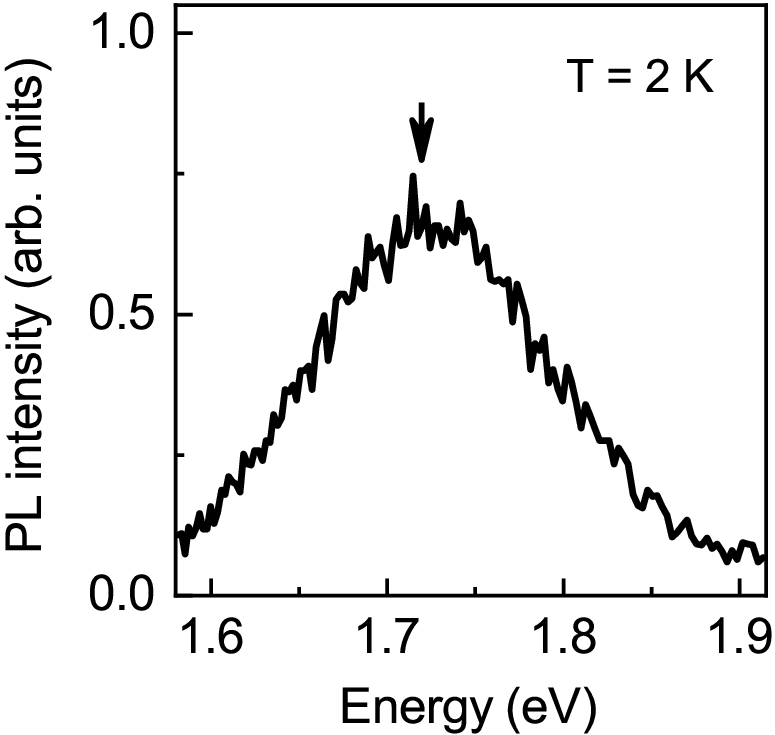}
\caption{PL spectrum of n-doped (In,Al)As/AlAs QDs. The arrow shows the energy 1.71~eV, at which the measurement of the circular polarization were performed, see Fig.~\ref{fig:B_dep}.}
  \label{FigS6}
\end{figure}

The PL spectrum at temperature $T=2$~K for this structure is shown in Fig.~\ref{FigS6}. The PL has the maximum at $1.71$~eV and the full width at half maximum approximately $140$~meV. The energy position of the PL band evidences the fact, that the conduction band minimum in these QDs belongs to the $X$-valley.

The circular polarization degree of PL  was measured at $1.71$~eV for the delay $t_{\rm d}=30$~$\mu$s with $t_{\rm g}=10$~$\mu$s. It is shown in
Fig.~\ref{fig:B_dep} by blue circles as a function of the magnetic field. As expected, there is no circular polarization at low magnetic fields. The polarization appears above $1$~T and saturates at $-0.92$ in the magnetic field of the order of $10$~T. This is the usual thermal spin polarization. Its negative sign confirms the fact that it is related with the negatively charged trions.

\subsection{E. Measurements in the Voight geometry}

Since the lifetime of the dark and bright neutral exciton states is very different, the PL dynamics at long time scales is related with the recombination of the dark excitons. A transverse magnetic field mixes the dark and bright exciton states~\cite{S_Ivchenko2018}, and accelerates the PL dynamics. The degree of mixing is determined by the ratio of the exchange splitting of the exciton states and the strength of the magnetic field.

\begin{figure}
  \centering
  \includegraphics[width=0.6\linewidth]{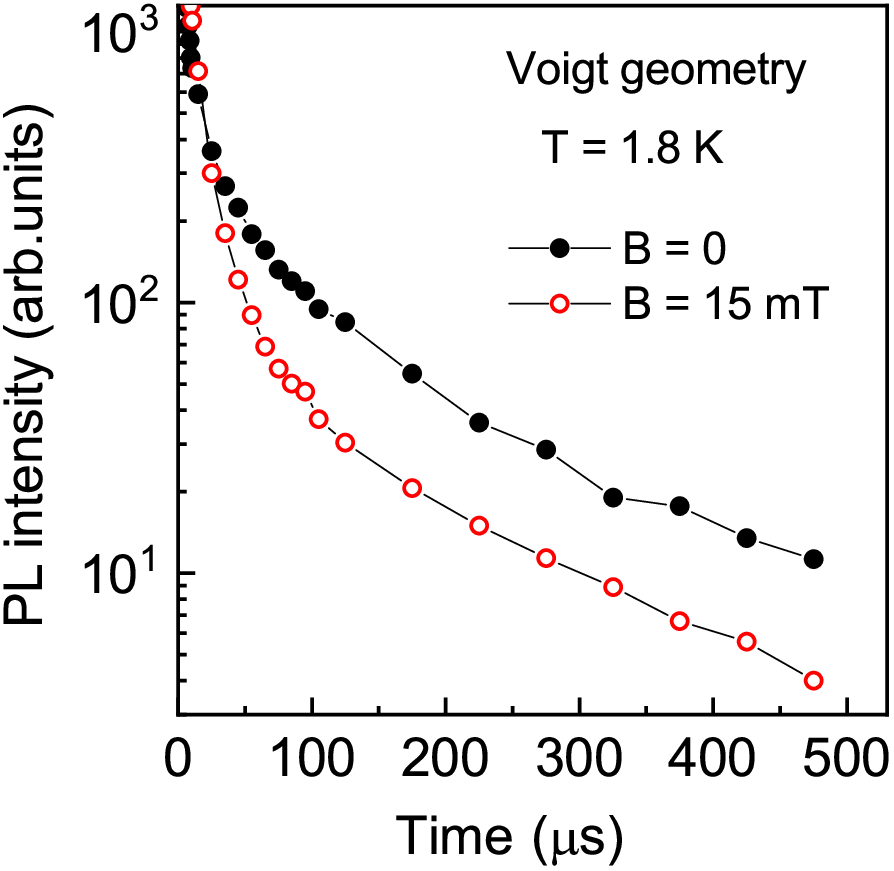}
\caption{Dynamics of the total PL of neutral excitons at the energy of $1.67$~eV in the Voigt geometry.}
  \label{fig:TZ}
\end{figure}

For the electrons in the $X$ valley, the exchange interaction with the hole in the $\Gamma$ valley is weak~\cite{S_birpikus_eng}. This results in the small splitting between the corresponding bright and dark exciton states. In order to evidence the fact that this splitting is small, we compare the PL dynamics in zero magnetic field and in the weak transverse magnetic field 15~mT (Voigt geometry). This comparison is shown in Fig.~\ref{fig:TZ}. The dynamics is not exponential due to a superposition of the multiple monoexponential contributions from the QDs with the different lifetimes~\cite{S_Shamirzaev84}. Nevertheless, one can see that the PL decay becomes faster in magnetic field due to the mixing of bright and dark exciton states. The PL decreases by the two orders of magnitude in $0.5$~ms in zero field and in $0.26$~ms in the transverse magnetic field.

\section{S3. Fit of experimental data}
\label{sec:fit}

\subsection{A. Dynamic spin polarization}

The curve in Fig.~\ref{fig3}(a) in the main text is calculated after the model of Eqs.~\eqref{eq:kinetic_av_S} with the parameters $\Delta_B=28$~mT, $B_{\rm exch}=6.6$~mT and $\tau_b=2~\mu$s. Note, that at the delay times shorter than $15~\mu$s, these parameters yield positive circular polarization degree $\sim0.05$. Experimentally it is not observed presumably because of the intense PL of bright excitons at these times.

To define the sign of the polarization we use the canonical  basis
for the circular polarized components of the electric
field~\cite{S_Varshalovich}. In this basis the polarization is
negative for the positive magnetic field along $z$-axis ($B_z>0$).
Indeed, from Fig.~\ref{fig:mechanism}(a) one can see, that in this
case the electron in exciton spin-flip time and dark exciton lifetime are
longer for the excitons with the heavy hole spin $J_z=-3/2$:
$\tau_{-2}>\tau_{+2}$, see Eq.~\eqref{eq:tau_pm} in the main text.
This is because the larger is the energy separation between the
states, the weaker is the nuclei assisted mixing of the exciton
states. As a result, at long times the $\sigma^-$ PL prevails for
$B_z>0$ resulting in $P_c<0$.

With temperature increase the spin relaxation of electron  in
exciton accelerates. Therefore, to describe the temperature
dependence shown in Fig.~\ref{fig4}(b) in the main text, we use
Eqs.~\eqref{eq:fast_all} (neglecting the dark exciton recombination
for simplicity). The decrease of the polarization degree at
temperatures above $7$~K evidences the abrupt acceleration of the
spin relaxation. To describe it quantitatively we assume
$\tau_s^e=\tau_s^{(0)}\exp(-k_BT/\Delta E)$ and use $\tau_s^{(0)}$
and $\Delta E$ as the fitting parameters. The magenta curve in
Fig.~\ref{fig4}(b) is calculated with the parameters
$\Delta_B=28$~mT, $B_{\rm exch}=6.6$~mT, $\tau_b=2~\mu$s,
$\tau_s^{(0)}=32~\mu$s and $\Delta E=0.2$~meV. This indicates, that
the spin relaxation at the elevated temperatures can be related with
a thermal population of some higher lying electron state, for
example, in a different $X$ valley.

\subsection{B. Thermal spin polarization}
\label{sec:T_fit}

In the previous analysis we considered four  exciton states. Here to
simplify the description of the thermal spin polarization we
consider only two phenomenological spin states split by $g_{\rm eff}
\mu_BB_z$ in external magnetic field. The thermal polarization in
the steady state is given by
\begin{equation}
  \label{eq:Pss}
  P_{\rm s.s.}=-P_0\tanh\left(\frac{g_{\rm eff} \mu_BB_z}{2k_BT}\right),
\end{equation}
where $0 \le P_0 \le 1$ is the depolarization factor. For the pulsed unpolarized excitation the time dependence of the polarization is described by
\begin{equation}
  P(t)=P_{\rm s.s.}\left[1-\exp{(-t/T_1)}\right],
  \label{eq:Pt}
\end{equation}
where $T_1$ is the longitudinal spin relaxation time. The best description of the experimental results is obtained if it scales as $T_1\propto B^3$. This can be related, for example, with fluctuations of the electric field leading to the spin relaxation due to the spin-orbit interaction~\cite{S_PhysRevB.71.165325}, or with the random changes of the hyperfine interaction caused by the piezoelectric electron-phonon interaction~\cite{S_Nazarov2002}. In this case we obtain
\begin{equation}
  \frac{t}{T_1}=\left(\frac{B_z}{B_0}\right)^3\coth\left(\frac{|g_{\rm eff}|\mu_BB_z}{2k_BT}\right),
  \label{eq:T1}
\end{equation}
where $B_0$ depends on the specific spin relaxation mechanism and scales as
\begin{equation}
  \label{eq:B0}
  B_0\propto t^{-1/3}
\end{equation}
with the delay after the pump pulse.

In Figs.~\ref{fig:B_dep} and~\ref{fig4}(a) in the main text the PL
polarization is shown in the large range of magnetic fields. For the
trions there is no dynamic spin polarization, so the corresponding
data can be fitted by Eq.~\eqref{eq:Pt}, as shown by the blue curves
in Figs.~\ref{fig:B_dep} and~\ref{fig4}(a) in the main text. The fit
parameters are $g_{\rm eff}=1.9$, $P_0=0.92$, and
$B_0=3.7$~T. For the neutral excitons at time delay $0.7~\mu$s there
is no dynamic spin polarization, so the PL polarization is again
described by Eq.~\eqref{eq:Pt}. The fit of experimental data is
shown in Fig.~\ref{fig:B_dep} by the green curve for the parameters
$g_{\rm eff}=-1.9$, $P_0=0.52$, and $B_0=13.5$~T. Finally, at the long delay
$70~\mu$s the dynamic and thermal polarization coexist, but take
place in the different ranges of magnetic fields, as shown in
Figs.~\ref{fig:B_dep} and~\ref{fig4}(a) in the main text. Therefore,
the polarization is described by the sum of Eqs.~\eqref{eq:simple_S}
and~\eqref{eq:Pt}. The fit is shown by the red curves, which are
calculated with the same parameters [including the rescaling of $B_0$
described by Eq.~\eqref{eq:B0}].

In strong magnetic field of $10$~T there is no dynamic  spin polarization, and the spin relaxation time is shorter, than the delay time, see Eq.~\eqref{eq:T1}. Therefore, the PL polarization shown in Fig.~\ref{fig4}(b) is described by Eq.~\eqref{eq:Pss} for the steady state with the parameters $g_{\rm eff}=-0.66$ and $P_0=0.69$.

\section{S4. Comparison of dynamic electron spin polarization with other mechanisms}

Here we consider alternative mechanisms of the electron spin polarization, which are documented via the PL circular polarization. We conclude, that they can not explain the presented experimental results. Therefore, the dynamic electron spin polarization is the only possible explanation.

For linearly polarized excitation, the conversion of the linear to circular polarization can take place in magnetic field~\cite{S_ivchenko91,S_PhysRevB.56.13405,S_dzhioev1998optical,S_Paillard2001}. We  have checked that the degree of circular polarization does not change with the rotation of the pump linear polarization, so this mechanism is irrelevant, as expected for the nonresonant excitation. The linear polarization of excitons could be also provided by the intrinsic anisotropy (birefringence) of the structure~\cite{S_PhysRevLett.83.3546,S_PhysRevB.61.R2421,S_PhysRevLett.88.257401,S_Platonov2002}. This does not happen in our case because: (i) The optical spin orientation is observed at quasi-resonant excitation in a very similar sample~\cite{S_Rautert99,S_PhysRevB.101.075412}. This means that there is no splitting between the linearly polarized exciton states. (ii) We have checked, that the linear polarization of the PL is almost zero. Therefore, the circular polarization can not be explained as a result of the conversion of the linear polarization to the circular one in magnetic field.

Also, it is known, that the thermal polarization of the exciton PL can be enhanced in the magnetic fields, which correspond to the anticrossings in the exciton fine structure levels~\cite{S_PhysRevB.26.827,S_Ivchenko2018}. A similar phenomenon is exploited for  the strain and magnetic field sensors based on the deep color centers in diamond~\cite{S_Awschalom1174,S_awschalom2018quantum} and silicon
carbide~\cite{S_PhysRevLett.112.187601,S_soltamov2019excitation}. In our model, the anticrossings of the levels take place in weak magnetic fields $B\sim B_{\rm exch}$, while the maximum circular polarization is reached at  higher fields $B\sim\Delta_B$. Note, that in the limit of the strong exchange interaction, $B_{\rm exch}\gg\Delta_B$, our model can also describe the circular polarization at the anticrossings of the fine structure levels. Therefore, the effect of dynamic electron spin polarization smoothly transforms to the thermal spin polarization with increase of the strength of the exchange interaction and external magnetic field.

For completeness, we note, that the PL circular polarization in small magnetic fields was reported for the phonon-assisted emission lines of excitons in Cu$_2$O~\cite{S_Gastev}. It was explained by the spin-dependent electron-phonon interaction in bulk crystals. This mechanism is also irrelevant for our experimental results.

\section{S5. Dynamic spin polarization of~resident electrons}

Here we consider a singly charged QD. The nonresonant optical excitation leads to the formation of a trion in the QD. In typical GaAs QDs, the lowest trion state is the singlet trion, where the two electrons are in the singlet state. There is no exchange interaction in the singlet trion state and, therefore, the dynamic electron spin polarization is not possible, which we confirmed experimentally, see Fig.~\ref{fig:B_dep}. Therefore, the dynamic spin polarization necessarily involves the triplet trion states. They can be either significantly populated due to the nonresonant excitation, or the triplet trion state can be the ground state in some structures. The latter is relevant, for example, for the QDs with the valley degeneracy of the conduction band according to the Hund's rules. This situation can be realized for some of the (In,Al)As QDs, in Si- and Ge-based QDs, as well as in the monolayers of transition metal dichalcogenides~\cite{S_PhysRevB.96.085302}. In the latter case, the exciton radiative recombination and electron-hole exchange interaction can be suppressed using the heterostructures, where electrons and holes reside in different monolayers~\cite{S_calman2018indirect}, while the hyperfine interaction for the localized excitons can be quite pronounced~\cite{S_MX2_Avdeev}.

\begin{figure}
  \centering
  \includegraphics[width=\linewidth]{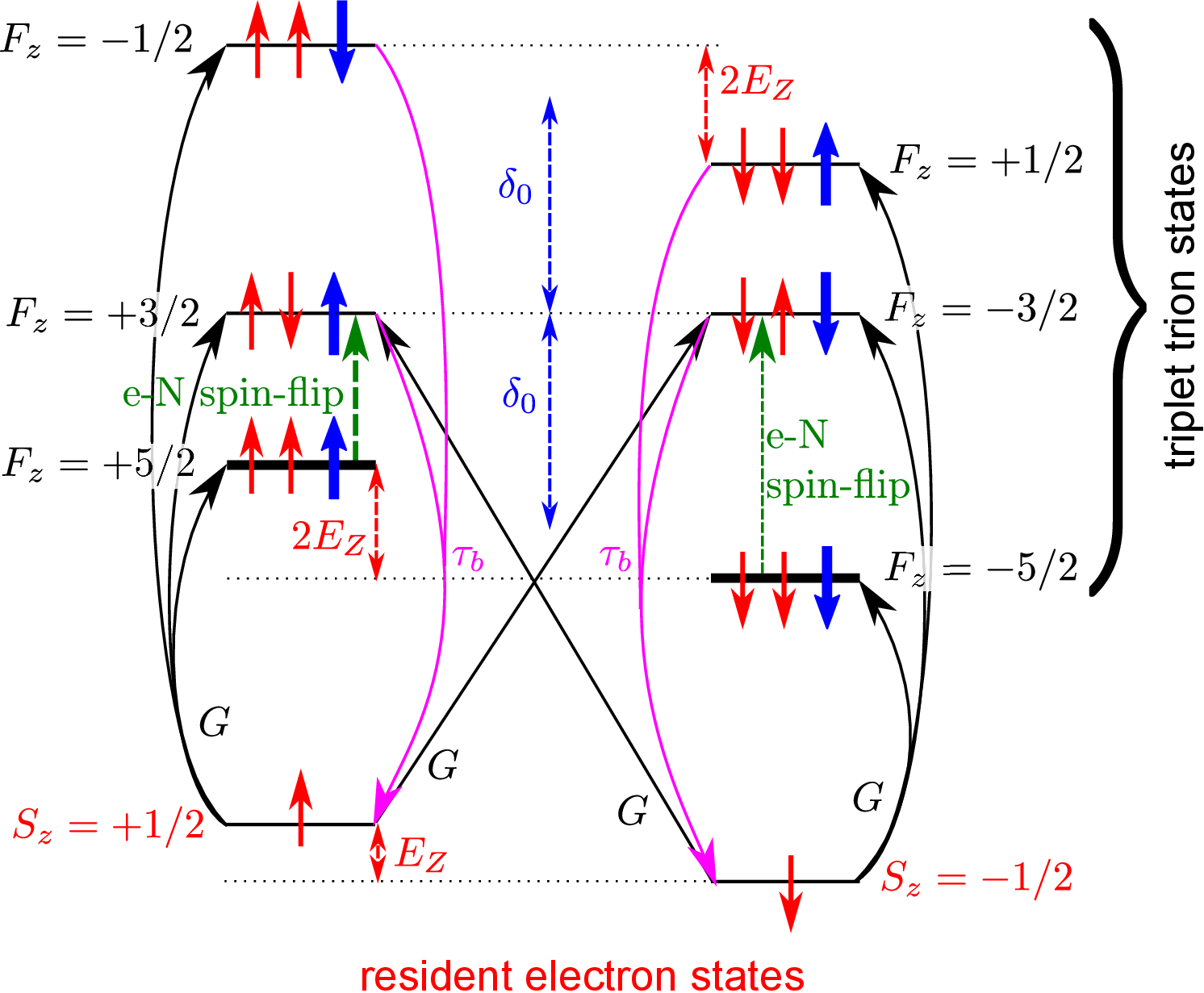}
  \caption{Mechanism of the resident electron dynamic spin polarization (for $g_h=0$) in a singly charged QD via the triplet trion state. The red and blue arrows denote the electron and heavy hole spins, respectively. The black arrows show nonresonant generation of the triplet trions with the rate $G$. The magenta arrows show the transitions corresponding to the radiative recombination. The mostly populated trion states are denoted by the thick lines. The green dashed arrows show the electron-nuclear spin flips, which lead to the dynamic electron spin polarization.}
  \label{fig:trions}
\end{figure}

Below we consider the triplet trion states only, which play the main role in dynamic spin polarization of resident electrons. We consider the six triplet trion states in a QD, as shown in Fig.~\ref{fig:trions}. Here the electron and heavy hole spins are denoted by the red and blue arrows, respectively, and it is assumed, that the two electrons belong to the two energy degenerate valleys of the conduction band and are always in the triplet state. The nonresonant and unpolarized optical excitation of the system leads to the generation of the electron hole pairs. The corresponding transitions are shown by the black arrows in Fig.~\ref{fig:trions}. It is assumed, that the photoexcited charge carriers quickly loose their spins and relax to one of the triplet trion states. The radiative trion recombination can bring the system back to the ground state, as shown by the magenta arrows.  We assume, that a hole can recombine with both electrons in the trion, for example, due to the interface mixing between the valleys.

Noteworthy, the radiative recombination is spin forbidden for the two lowest trion states where all spins of the three charge carriers are parallel. These states are shown by the thicker lines in Fig.~\ref{fig:trions} and have the total angular momentum $F_z=\pm5/2$. These states are preferentially populated in the steady state. The trion in these states can recombine due to the electron-nuclear spin flips only, which mix these states with the trion states corresponding to $F_z=\pm3/2$, as shown by the green arrows. In external magnetic field, the energy difference in the pairs of stares with $F_z=+5/2,+3/2$ and $F_z=-5/2,-3/2$ is different by $2E_Z$. If $B_z>0$ and $g_e>0$, then the trion with $F_z=+5/2$ recombines faster than the trion with $F_z=-5/2$. Qualitatively, this results in the faster population of the ground electron state with $S_z=+1/2$, e.g. in the dynamic spin polarization of a resident electron. The detailed quantitative analysis of the dynamic polarization of the resident electrons will be reported elsewhere.


%
\end{document}